\documentclass{article}
\usepackage{amssymb}
\usepackage{amsmath}
\usepackage{amsfonts}
\usepackage{authblk}

\begin{document}
\title{Integrability of the geodesic flow on the resolved conifolds over 
Sasaki-Einstein space $T^{1,1}$}

\author{Mihai Visinescu\thanks{mvisin@theory.nipne.ro}}

\affil{Department of Theoretical Physics,

National Institute for Physics and Nuclear Engineering,

Magurele, P.O.Box M.G.-6, Romania}

\date{}
\maketitle

\begin{abstract}

Methods of Hamiltonian dynamics are applied to study the geodesic flow on the resolved
conifolds over Sasaki-Einstein space $T^{1,1}$. We construct explicitly the constants 
of motion and prove complete integrability of geodesics in the five-dimensional 
Sasaki-Einstein space $T^{1,1}$ and its Calabi-Yau metric cone. The singularity at 
the apex of the metric cone can be smoothed out in two different ways. Using
the small resolution the geodesic motion on the resolved conifold remains completely
integrable. Instead, in the case of the deformation of the conifold the complete 
integrability is lost.

\end{abstract}






\section{Introduction}

There has been considerable interest recently in contact geometry in connection with 
some modern developments in mathematics and theoretical physics \cite{B-G-2008,Sp}.
Sasaki-Einstein geometry is considered important in studies of
consistent string compactification and in the context of AdS/CFT correspondence
\cite{JMM,GMSW}.

The symmetries of Sasaki-Einstein spaces play a significant role in connection with
the study of integrability properties of geodesic motions and separation of variables 
of the classical Hamilton-Jacobi or quantum Klein-Gordon, Dirac equations.

The homogeneous toric Sasaki-Einstein on $S^2 \times S^3$ is usually referred to as
$T^{1,1}$ and was considered as the first example of toric Sasaki-Einstein/qui\-ver duality 
\cite{KW}. The $AdS \times T^{1,1}$ is the first example of a supersymmetric holographic
theory based on a compact manifold which is not locally $S^5$.

In a recent paper \cite{BV} it has been constructed the complete set of constants of 
motion for Sasaki-Einstein spaces $T^{1,1}$ and $Y^{p,q}$ proving the complete 
integrability of geodesic flows. Moreover it was used the action-angle formalism to 
study thoroughly the complete integrability of these systems \cite{MVEPJC,MVPTEP}.

In this paper we are concerned with integrability of geodesic motions in Sasaki-Einstein
space $T^{1,1}$ and its Calabi-Yau metric cone. In general the metric cone of a 
Sasaki-Einstein manifold is singular at the apex of the cone. On the other hand there are
promising generalizations of the original AdS/CFT correspondence by considering $D3$-branes
on conifold singularities \cite{KW,MP}.

Regarding the metric cone over $T^{1,1}$, the singularity at its apex can be 
smoothed out in two different ways \cite{C-O}. On can substitute the apex by an $S^3$ 
(\emph{deformation}) or by an $S^2$ (\emph{small resolution}). We find out
that the geodesic motions are completely integrable in the case of the metric cone
as well as its small resolution of the apex singularity. In opposition to these cases,
the geodesic flow on the deformed metric cone is not completely integrable.

We proceed to use the action-angle approach for these integrable systems with a caveat 
that the radial motion is unbounded. One may separate the radial from the angular degrees 
of freedom. The compact phase space corresponding to the angular degrees of freedom can 
be formulated in terms of action-angle variables and, finally, add a radial part to it.

The paper is organized as follows. In the next Section we briefly describe the complete 
integrability of the geodesic motion in $T^{1,1}$ space. In Section 3 we proceed to  
investigate the integrability of the geodesic flow on the metric cone over $T^{1,1}$.
In Section 4 we extend the study of the integrability to the resolved conifolds in both 
known procedures of smoothing out the singularity at the apex of the conifold. The paper
ends with conclusions in Section 5.

\section{Complete integrability on $T^{1,1}$ space} 

The homogeneous Sasaki-Einstein space $T^{1,1}$ has topology $S^2 \times S^3$ and 
its isometries form the group $SU(2) \times SU(2) \times U(1)$. The metric on this 
space is particularly simple \cite{C-O,M-S}:
\[
ds^2_{T^{1,1}} =  
\frac16 (d \theta^2_1 + \sin^2 \theta_1 d \phi^2_1 +
d \theta^2_2 + \sin^2 \theta_2 d \phi^2_2) 
+\frac19 (d \psi + \cos \theta_1 d \phi_1 + \cos \theta_2 d \phi_2)^2 \,.
\]
Here $\theta_i\,,\, \phi_i\, , i=1,2$ are the usual polar and axial coordinates on two
round two-sphere and $\psi \in [0, 4 \pi)$ parametrizes the $U(1)$ bundle over 
$S^2 \times S^2$. 

The Hamiltonian describing the geodesic flow is
\begin{equation}\label{Ham}
H = \frac12 g^{ij}\, p_i p_j \,, 
\end{equation}
where $g^{ij}$ is the inverse metric of $T^{1,1}$ space and $p_i = g_{ij} \,\dot{x}^j$ 
are the conjugate momenta to the coordinates $(\theta_1,\theta_2,\phi_1,\phi_2,\psi)$. 

Let us denote by $(p_{\theta_1},p_{\theta_2},p_{\phi_1},p_{\phi_2},p_\psi)$ these
conjugate momenta:
\begin{equation}\label{momT}
\begin{split}
p_{\theta_1} &= \frac16 \dot{\theta}_1\,,\\
p_{\theta_2} &= \frac16 \dot{\theta}_2\,, \\
p_{\phi_1} &= \frac16 \sin^2\theta_1 \,\dot{\phi}_1 + 
\frac19 \cos\theta_1(\cos\theta_1 \,\dot{\phi}_1
+ \cos\theta_2\,\dot{\phi}_2 + \dot{\psi})\,,\\
p_{\phi_2} &= \frac16 \sin^2\theta_2 \,\dot{\phi}_2 + 
\frac19 \cos\theta_2(\cos\theta_1 \,\dot{\phi}_1
+ \cos\theta_2\,\dot{\phi}_2 + \dot{\psi})\,,\\
p_{\psi} &= \frac19 ( \cos\theta_1 \,\dot{\phi}_1 + 
\cos\theta_2 \,\dot{\phi}_2 + \dot{\psi})\,.
\end{split}
\end{equation}

In terms of them, the Hamiltonian \eqref{Ham} has the form:
\begin{equation}\label{HamT}
\begin{split}
H=&
3 \left[ p^2_{\theta_1} +  p^2_{\theta_2} +
\frac{1}{\sin^2\theta_1}( p_{\phi_1} - \cos\theta_1 \, p_{\psi})^2 +
\frac{1}{\sin^2\theta_2}( p_{\phi_2} - \cos\theta_2 \, p_{\psi})^2 \right] \\
&+ \frac92 \, p^2_{\psi} \,.
\end{split}
\end{equation}

The Hamiltonian is cyclic in $\phi_1 ,\phi_2 ,\psi$ and consequently their conjugate 
momenta $p_{\phi_1},p_{\phi_2} , p_{\psi}$ are conserved. Besides, two total $SU(2)$ 
angular momenta are also conserved:
\[
\begin{split}
\mathbf{j}_1^{2} &= p_{\theta_1}^2 + \frac{1}{\sin^2\theta_1}( p_{\phi_1}
- \cos\theta_1 \,p_{\psi})^2  +p^2_{\psi} \,,\\
\mathbf{j}_2^{2} &= p_{\theta_2}^2 + \frac{1}{\sin^2\theta_2}( p_{\phi_2}
- \cos\theta_2 \, p_{\psi})^2  +p^2_{\psi} \,.
\end{split}
\]

Other constants of motion can be constructed using the St\"{a}ckel-Killing tensors
of the $T^{1,1}$ space \cite{BV,S-V-V}. In spite of the presence of a multitude of 
conserved quantities, the number of functionally independent constants of motion is 
five. This implies the complete integrability of geodesic motions in $T^{1,1}$ which
allows us to solve the Hamilton-Jacobi equation by separation of variables
and construct the action-angle variables \cite{MVEPJC}.

\section{Complete integrability on the metric cone}

It is well known that a compact Riemannian manifold $(M, g_M)$ is Sasakian if and only 
if its metric cone 
\[
\bigl(C(M),\, g_{C(M)}\bigr) = \bigl(\mathbb{R}_+ \times M\,,dr^2 + r^2 g_{M}\bigr) \,, 
\]
is K\"{a}hler \cite{B-G-2008}. Here $r \in (0,\infty)$ may be considered as a coordinate 
on the positive real line $\mathbb{R}_+$. Moreover, if the Sasakian manifold is Einstein,
the metric cone is Ricci-flat and K\"{a}hler, i.e., Calabi-Yau.

In connection with the Sasaki-Einstein space $T^{1,1}$, the conifold metric (mc) is
\begin{equation}\label{mc}
 ds^2_{mc} = dr^2 + r^2\left(\frac16 \sum_{i=1}^2\left(d \theta^2_i 
 + \sin^2 \theta_i d \phi^2_i\right) 
+\frac19 \left(d \psi + \sum_{i=1}^2\cos \theta_i d \phi_i\right)^2 \right)\,.
\end{equation}

On the metric cone \eqref{mc} the geodesic flow is described by the Hamiltonian:
\begin{equation}\label{Hammc}
H_{C(T^{1,1})} = \frac12 \, p_r^2 + \frac{1}{r^2}\, \tilde{H} \,,
\end{equation}
where the radial momentum is
\[
p_r = \dot{r}\,.
\]

The Hamiltonian $\tilde{H}$ has a similar structure as in \eqref{HamT}, but constructed 
with momenta $\tilde{p}_i$ related to momenta $p_i$ \eqref{momT} by
\[
\tilde{p}_i = r^2 g_{ij} \,\dot{x}^j = r^2 \, p_i \,. 
\]

It is not difficult to observe that the radial dynamics is independent of the 
dynamics of the base manifold $T^{1,1}$ and $\tilde{H}$ is a constant of motion.
The Hamilton equations of motion for $\tilde{H}$ on $T^{1,1}$ have the standard 
form in terms of a new time variable $\tilde{t}$ given by \cite{FMW}
\[
\frac{d t}{d \tilde{t}} = r^2\,. 
\]

Concerning the constant of motions, they are the conjugate momenta
$(\tilde{p}_{\phi_1},\\ \tilde{p}_{\phi_2},\tilde{p}_\psi)$ associated with the cyclic
coordinates $(\phi_1, \phi_2, \psi)$ and two total $SU(2)$ momenta
\[
\tilde{\mathbf{j}}_i^{2} = r^4 \,\mathbf{j}_i^{2}\quad , \quad i=1,2 \,.
\]
Together with the Hamiltonian $H_{C(T^{1,1})}$, they ensure the complete 
integrability of the geodesic flow on the metric cone.

Considering a particular level set $E$ of $H_{C(T^{1,1})}$,
we get for the radial motion
\[
p^2_r = \dot{r}^2 = 2 E - \frac2{r^2} \, \tilde{H}\,.
\]
The turning point of the radial motion is determined by
\begin{equation}\label{tp}
\dot{r} =0 \quad \Longrightarrow \quad r_* = \sqrt{\frac{\tilde{H}}{E}} \,.
\end{equation}

Projecting the geodesic curves onto the base manifold $T^{1,1}$ we can evaluate 
the total distance transversed in the Sasaki-Einstein space between the limiting 
points as $t\rightarrow -\infty$ and $t\rightarrow +\infty$ \cite{FMW}
\[
d= \sqrt{2\tilde{H}}\int^{\infty}_{-\infty} \frac {dt}{r^2_* + 2E \,t^2} 
= \pi \,.
\]

To proceed with the analysis of the integrability of geodesic motions on the metric 
cone we must be aware that the radial motion is unbounded and consequently  the Hamiltonian
\eqref{Hammc} does not admit a formulation in terms of action-angle variables. The key idea
\cite{HKLN,LNY,HLNSY} is to split the mechanical system into a ``radial'' and an 
``angular'' part. The angular part is a compact subsystem spanned by the set of variables 
$q=(\theta_1, \theta_2, \phi_1, \phi_2, \psi)$ which can be formulated in terms of 
action-angle variables $(I_i, \Phi^0_i)\,,\, i=\theta_1, \theta_2, \phi_1, \phi_2, \psi$.
By adding the radial part $r \in (0,\infty)$, the Hamiltonian \eqref{Hammc} can be put 
in the form
\[
H_{C(T^{1,1})} = \frac12 \, p_r^2 + \frac{1}{r^2}\, \tilde{H}(I_i) \,.
\]

Concerning the action-angle variables $(I_i, \Phi^0_i)$ corresponding to the compact angular 
subsystem, they can be determined by a standard technique. To fulfill this task, let us
consider a particular level set $E$ of the energy. Using complete separability, 
Hamilton's principal function can be written in the form \cite{GPS}:
\begin{equation}\label{Hpf}
\begin{split}
S(r,q,\alpha, t)  = &~S_0(r,q,\alpha) - Et  = S_r(r,\alpha) + \tilde{S}_0(q,\alpha) - Et\\
=&~ S_r(r,\alpha) +
\sum_{j=1,2} S_{\theta_j}(\theta_j,\alpha) + \sum_{j=1,2} S _{\phi_j}(\phi_j,\alpha) 
+ S _{\psi}(\psi,\alpha)-Et\,,
\end{split}
\end{equation}
where $\alpha$ is a set of constants of integration.

The Hamilton-Jacobi equation is
\begin{equation}\label{HJe}
\begin{split}
E = &~\frac12 \left( \frac{\partial S_r}{\partial r}\right)^2 + \frac{3}{r^2} \sum_{i=1,2}
\left\{\left( \frac{\partial S_{\theta_i}}{\partial \theta_i}\right)^2
+ \frac1{\sin^2\theta_i}\left[\left( \frac{\partial S_{\phi_i}}{\partial \phi_i}\right)
- \cos\theta_i \left( \frac{\partial S_{\psi}}{\partial \psi}\right)\right]^2 \right\}\\
&~+ \frac9{2r^2} \left( \frac{\partial S_{\psi}}{\partial \psi}\right)^2 \,.
\end{split}
\end{equation}

Since the variables $(\phi_1,\phi_2,\psi)$ are cyclic, we have
\[
\begin{split}
S_{\phi_1} =& ~\tilde{p}_{\phi_1} \cdot \phi_1 = \alpha_{\phi_1} \cdot \phi_1 \,,\\
S_{\phi_2} =& ~\tilde{p}_{\phi_2} \cdot \phi_2 = \alpha_{\phi_2} \cdot \phi_2 \,,\\
S_{\psi} =& ~\tilde{p}_{\psi} \cdot \psi =  \alpha_{\psi} \cdot \psi\,,
\end{split}
\]
where $\alpha_{\phi_1}, \alpha_{\phi_2}, \alpha_{\psi}$ are constants of integration.
The corresponding action variable are:
\[
\begin{split}
I_{\phi_1} = &~\frac1{2\pi} \oint \frac{\partial S_{\phi_1}}{\partial \phi_1} d\phi_1 
= \alpha_{\phi_1}\,,\\
I_{\phi_2} = &~\frac1{2\pi} \oint \frac{\partial S_{\phi_12}}{\partial \phi_2} d\phi_2 
= \alpha_{\phi_2}\,,\\
I_{\psi} = &~\frac1{4\pi} \oint \frac{\partial S_{\psi}}{\partial \psi} d\psi 
= \alpha_{\psi}\,.
\end{split}
\]

Next we deal with the coordinates $\theta_i\,,\,i=1,2$. From Hamilton-Jacobi equation 
\eqref{HJe} we get
\[
\left( \frac{\partial S_{\theta_i}}{\partial \theta_i} \right)^2
+ \frac1{\sin^2\theta_i}\left( \alpha_{\phi_i} - \alpha_\psi \cos\theta_i\right)^2=
\alpha^2_{\theta_i} \quad ,\quad i=1,2\,,
\]
where $\alpha_{\theta_i} \,,\, i=1,2$ are constants.

The corresponding action variables  $I_{\theta_i}\,,\, i=1,2$, are
\[
I_{\theta_i} = \frac1{2\pi} \oint \left( \alpha^2_{\theta_i} - 
\frac{(\alpha_{\phi_i} - \alpha_\psi \cos\theta_i)^2}{\sin^2\theta_i}\right)^{\frac12} 
d \theta_i \quad , \quad i=1,2\,.
\]

The limits of integrations are defined by the roots $\theta_{i-}$ and $\theta_{i+}$
of the expressions in the square root parenthesis and a complete cycle of $\theta_i$
involves going from $\theta_{i-}$ to $\theta_{i+}$ and back to $\theta_{i-}$.
An efficient technique for evaluating $I_{\theta_i}$  is to extend $\theta_i$ to a complex
variable $z_i$ and interpret the integral as a closed contour integral in the $z_i$ plane 
\cite{MVEPJC,MVPTEP}. At the end, we get
\[
I_{\theta_i} = \left( \alpha^2_{\theta_i} +   \alpha^2_{\psi} \right)^{\frac12}
- \alpha_{\phi_i} \quad , \quad i=1,2 \,.
\]

We note that each $S_i\,,\, i=(\theta_1, \theta_2, \phi_1, \phi_2, \psi)$ depends on all 
action variables $I_i$. Then the angle variables are
\[
\Phi^0_i = \frac{\partial \tilde{S}_0}{\partial I_i} \quad,\quad  
i=(\theta_1, \theta_2, \phi_1, \phi_2, \psi)\,.
\]
Their explicit expressions are quite involved and are not produced here, but they can be 
found in \cite{MVEPJC,MVPTEP}.

Finally, let us consider the radial part of the Hamilton's principal function. From 
Hamilton-Jacobi equation \eqref{HJe} we have
\[
\left(\frac{\partial S_r}{\partial r} \right)^2 = 2E - \frac6{r^2}
(\alpha^2_{\theta_1} + \alpha^2_{\theta_2} +\frac32 \alpha^2_{\psi})\,,
\]
and
\[
S_r(r,\alpha) = \int^r d r' \left(2E - \frac6{r'^2}
(\alpha^2_{\theta_1} + \alpha^2_{\theta_2} + \frac32 \alpha^2_{\psi})\right)^{\frac12}\,.
\]

Thus, the Hamilton's principal function \eqref{Hpf} is 
\begin{equation}\label{Hpff}
S(E,I_i,r,\Phi^0_i) = \sqrt{2} \int^r d r' \sqrt{E - \frac{\tilde{H}(I_i)}{r'^2}}
+ \sum_i I_i \Phi^0_i \,,
\end{equation}
where the sum extends over the action-angle variables 
$i=(\theta_1, \theta_2, \phi_1, \phi_2, \psi)$.

The integral in \eqref{Hpff} can be evaluated and, eventually, we get
\[
S_r(r,I_i) = \sqrt{2E} \left( \sqrt{r^2 - r^2_*} - 
r_* \arctan{\sqrt{\frac{r^2- r^2_*}{r^2_*}}}\right)\, 
\]
where the turning point $r_*$ \eqref{tp}, in terms of action variables $I_i$, is
\[
r^2_*= \frac3E \left[ (I_{\theta_1} + I_{\phi_1})^2 + (I_{\theta_1} + I_{\phi_1})^2 
- \frac12 I_\psi^2\right]\,. 
\]

\section{Integrability of the resolved conifolds}

The metric cone over Sasaki-Einstein space $T^{1,1}$ has a singularity at its apex
which can be repaired in two different ways \cite{C-O}. The first is represented by a 
\emph{deformation} having the effect of replacing the node by a sphere $S^3$. The second 
possibility consists in a replacement of the node by an $S^2$ (\emph{small resolution}).

The metric cone \eqref{mc} associated with Sasaki-Einstein space $T^{1,1}$ is described 
by the following equation in four complex variables
\begin{equation}\label{quadric}
\sum_{a=1}^4 w_a^2 =0\,.
\end{equation}
The equation of the quadric can be rewritten using a matrix $\mathcal{W}$
defined by
\begin{equation}\label{W}
\mathcal{W} = \frac1{\sqrt 2}\,  \sigma^a w_a =
\frac1{\sqrt 2}\, 
\left( \begin{array}{cc}
w_3 + i w_4 \,&\, w_1 - i w_2\\
w_1 +i w_2 \,& \,-w_3 +i w_4
\end{array} \right) 
\equiv 
\left( \begin{array}{cc}
X \,&\, U\\
V \,&\, Y
\end{array} \right) \,,
\end{equation} 
where $\sigma^a$ are the Pauli matrices for $a=1,2,3$ and $\sigma^4$ is $i$ times 
the unit matrix. The radial coordinate is defined by
\[
r^2 = \text{tr} \, ( \mathcal{W}^\dagger \mathcal{W})\,.
\]

In terms of the matrix $\mathcal{W}$, equation \eqref{quadric} can be written as
\begin{equation}\label{detW}
\text{det} \, \mathcal{W}  =0 \quad , \quad \text{i.e.} \quad XY -UV =0 \,.
\end{equation}

\subsection{Small resolution}

The small resolution is realized replacing \eqref{detW} by the pair of equations 
\cite{C-O}:
\[
\left( \begin{array}{cc}
X \,&\, U\\
V \,&\, Y
\end{array} \right) 
\left( \begin{array}{c}
\lambda_1 \\
\lambda_2
\end{array} \right) = 0\,,
\]
in which $(\lambda_1, \lambda_2) \in \mathbb{CP}^1$ are not both zero. 
Thus, in the region where $\lambda_1 \neq 0$, the pair $(\lambda_1, \lambda_2)$ is 
uniquely characterized by the coordinate $\lambda = \lambda_2 / \lambda_1$, while 
in the region where $\lambda_2 \neq 0$, the pair $(\lambda_1, \lambda_2)$ is 
described by the coordinate $\mu = \lambda_1 / \lambda_2$.

It turns out to be convenient to introduce a new radial coordinate
\[
\rho^2 \equiv \frac32 \, \gamma \,,
\]
where the function $\gamma$ is given by the equation
\[
\gamma^3 + 6\, a^2 \,\gamma^2 - r^4 =0\,,
\]
with $a$ the ``resolution'' parameter. It represents the radius of the sphere 
$S^2$ which replaces the point singularity at $r^2=0$.

Eventually, the metric of the resolved conifold (rc) can be written as \cite{PZT}
\begin{equation}\label{mrc}
\begin{split}
d s^2_{rc}= & \,\kappa^{-1}(\rho)\, d\rho^2 + \frac19 \kappa(\rho) \rho^2
(d \psi + \cos \theta_1 \, d \phi_1 + \cos \theta_2 \, d \phi_2)^2\\
&+\frac16 \rho^2 (d \theta^2_1 + \sin^2 \theta_1 \, d \phi^2_1)
+\frac16 (\rho^2 + 6 \, a^2)
(d \theta^2_2 + \sin^2 \theta_2 \, d \phi^2_2) \,,
\end{split}
\end{equation}
where
\[
\kappa(\rho) \equiv \frac{\rho^2 + 9 \,a^2}{\rho^2 + 6 \,a^2} \,.
\]

The resolved conifold metric is Ricci flat and has an explicit $SU(2) \times SU(2)$
invariant form. When the resolution parameter $a$ goes to zero or when
$\rho \rightarrow \infty$, the resolved conifold metric reduces to the standard 
conifold metric  $g_{C(T^{1,1})}$. In fact the parameter $a$ introduces an asymmetry 
between the two sphere.

The conjugate momenta $(P_\rho,P_{\theta_1},P_{\theta_2},P_{\phi_1},P_{\phi_2},P_\psi)$
corresponding to the coordinates $(\rho,\theta_1,\theta_2,\phi_1,\phi_2,\psi)$ are:
\[
\begin{split}
P_\rho&= \kappa^{-1}(\rho) \dot{\rho}\,,\\
P_{\theta_1} &= \frac16 \rho^2 \dot{\theta}_1\,,\\
P_{\theta_2} &= \frac16 (\rho^2 + 6\, a^2)\dot{\theta}_2\,, \\
P_{\phi_1} &= \frac16 \rho^2 \sin^2\theta_1 \,\dot{\phi}_1 + 
\frac19 \kappa(\rho) \rho^2 \cos\theta_1 
(\cos\theta_1 \,\dot{\phi}_1 +  \cos\theta_2\,\dot{\phi}_2 +\dot{\psi})\,,\\
P_{\phi_2} &= \frac16 (\rho^2 + 6\, a^2)\sin^2\theta_2 \,\dot{\phi}_2 + 
\frac19 \kappa(\rho) \rho^2 \cos\theta_2
(\cos\theta_1 \,\dot{\phi}_1 +  \cos\theta_2\,\dot{\phi}_2 + \dot{\psi})\,,\\
P_{\psi} &= \frac19 \kappa(\rho) \rho^2 (\cos\theta_1 \,\dot{\phi}_1 + 
\cos\theta_2 \,\dot{\phi}_2 +\dot{\psi} )\,.
\end{split}
\]

In terms of them, the Hamiltonian for the geodesic flow on rc is
\[
\begin{split}
H_{rc} =&~  \frac{\kappa(\rho)}2 P_\rho^2 + \frac92\,\frac1{\kappa(\rho) \, \rho^2}
P_\psi^2 + \frac3{\rho^2} \, P_{\theta_1}^2 + \frac3{\rho^2 + 6\, a^2}\, P_{\theta_2}^2 \\
& + \frac3{\rho^2 \sin^2\theta_1} \,(P_{\phi_1} - \cos\theta_1 \,P_\psi)^2 \\
& +\frac3{(\rho^2 + 6 \,a^2) \,\sin^2\theta_2} \,(P_{\phi_2} - \cos\theta_2\, P_\psi)^2\,.
\end{split}
\]

We observe that ($\phi_1,\phi_2, \psi$) are still cyclic coordinates and, accordingly, 
momenta $P_{\phi_1},P_{\phi_2}, P_\psi$ are conserved. Taking into account the 
$SU(2)\times SU(2)$ symmetry of the metric \eqref{mrc}, the total angular momenta
\[
\begin{split}
\mathbf{J}_1^{2} &= P_{\theta_1}^2 + \frac{1}{\sin^2\theta_1}( P_{\phi_1}
- \cos\theta_1 \,P_{\psi})^2  + P^2_{\psi} 
=   \rho^4 \,\mathbf{j}_1^{2}\,,\\
\mathbf{J}_2^{2} &= P_{\theta_2}^2 + \frac{1}{\sin^2\theta_2}( P_{\phi_2}
- \cos\theta_2 \, P_{\psi})^2  + P^2_{\psi} 
= (\rho^2 + 6 a^2)^2\,\mathbf{j}_2^{2}\,,
\end{split}
\]
are also conserved. The set of conserved quantities 
($H_{rc},P_{\phi_1},P_{\phi_2}, P_\psi,\mathbf{J}_1^{2},\mathbf{J}_2^{2}$)
ensures the complete integrability of geodesic motions on the resolved conifold.
As it is expected, for $a=0$ we recover the state of integrability on the standard
metric cone of the Sasaki-Einstein space $T^{1,1}$.

In what follows, as in the case of the metric cone, we separate the radial from the 
angular degrees of freedom. We shall look for the Hamilton's principal function 
in the form:
\[
S(\rho,q,\alpha, t) 
= S_\rho(\rho,\alpha) +
\sum_{j=1,2} S_{\theta_j}(\theta_j,\alpha) + \sum_{j=1,2} S _{\phi_j}(\phi_j,\alpha) 
+ S _{\psi}(\psi,\alpha)-Et\,.
\]

The Hamilton-Jacobi equation becomes:
\[
\begin{split}
E = &~\frac12 \kappa (\rho)\left( \frac{\partial S_\rho}{\partial \rho}\right)^2 + \frac{3}{\rho^2} 
\left( \frac{\partial S_{\theta_1}}{\partial \theta_1}\right)^2 + \frac{3}{\rho^2 + 6 a^2} 
\left( \frac{\partial S_{\theta_2}}{\partial \theta_2}\right)^2\\
&~+ \frac3{\rho^2\sin^2\theta_1}\left[\left( \frac{\partial S_{\phi_1}}{\partial \phi_1}\right)
- \cos\theta_1 \left( \frac{\partial S_{\psi}}{\partial \psi}\right)\right]^2 \\
&~+ \frac3{(\rho^2+ 6a^2)\sin^2\theta_2}\left[\left( \frac{\partial S_{\phi_2}}{\partial \phi_2}\right)
- \cos\theta_2 \left( \frac{\partial S_{\psi}}{\partial \psi}\right)\right]^2 \\
&~+ \frac9{2\kappa(\rho) \rho^2} \left( \frac{\partial S_{\psi}}{\partial \psi}\right)^2 \,.
\end{split}
\]

As before, $\phi_1,\phi_2,\psi$ are cyclic coordinates and the evaluation of $S_{\phi_1},
S_{\phi_2},S_{\psi}$ proceeds as in the previous Section. Fortunately, the evaluation of 
$S_{\theta_1}$ and $S_{\theta_2}$ is again as above.

Concerning the radial part of the Hamilton's principal function we get a more intricate equation:
\[
\begin{split}
E = &~\frac12 \kappa (\rho)\left( \frac{\partial S_\rho}{\partial \rho}\right)^2 +
\frac9{2\kappa(\rho) \rho^2} I^2_\psi + \frac3{\rho^2} \left[ (I_{\theta_1} + 
I_{\phi_1})^2 - I^2_\psi \right] \\
&~+\frac3{(\rho^2 + 6 a^2)} \left[ (I_{\theta_2} + 
I_{\phi_2})^2 - I^2_\psi \right]\,.
\end{split}
\]

This equation can be integrated, but the result is not at all illuminating to be produced
here. Anyway, we remark the asymmetry between the contribution to the Hamiltonian of the 
action variables associated with the motions on the two sphere $S^2$. This contrasts with 
the situation of the geodesic flow on the metric cone of $T^{1,1}$. 
Moreover, in the case of $T^{1,1}$, the study of integrability points out that two pairs 
of frequencies of the geodesic motion are resonant and the Hamiltonian involves a reduced 
number of action variables.

\subsection{Deformation}

The deformation of the conifold consists in replacing the apex by an $S^3$ which is
achieved by another modification of equation \eqref{quadric}. The metric cone is deformed
to a smooth manifold described by the equation:
\[
\sum_{a=1}^4 w_a^2 =\epsilon^2 \,,
\]
where $\epsilon$ is the ``deformation'' parameter. In terms of the matrix $\mathcal{W}$ 
\eqref{W}, equation \eqref{detW} becomes
\[
det \, \mathcal{W} = - \frac12 \epsilon^2\,.
\]

On setting the new radial coordinate
\[
r^2 = \epsilon^2 \cosh \tau\,,
\]
the deformed conifold (dc) metric is \cite{C-O,OY,MT}:
\begin{equation}\label{metricdc}
ds^2_{dc} = \frac12 \epsilon^{\frac43} K(\tau)\,\left(\frac1{3K^3(\tau)}
\,(d\tau^2 + ds^2_1) + \frac{\cosh \tau}2 \,ds^2_2 + \frac12 \,ds^2_3
\right)\,,
\end{equation}
where 
\[
K(\tau) =\frac{(\sinh 2\tau - 2\tau)^\frac13}{2^\frac13 \,\sinh \tau}\,,
\]
and
\[
\begin{split}
ds^2_1 =&~ (d\psi + \cos \theta_1 \,d\phi_1 + \cos \theta_2\, d\phi_2)^2\,,\\
ds^2_2 =&~ d\theta_1^2 + d\theta_2^2 + \sin^2\theta_1 \, d\phi_1^2 + 
\sin^2\theta_2 \, d\phi_2^2\,,\\
ds^2_3 =&~ 2 \bigl( \sin\psi (d\phi_1 \, d\theta_2 \, \sin\theta_1
+d\phi_2 \, d\theta_1 \, \sin\theta_2) \bigr.\\
&\bigl. ~ + \cos\psi(d\theta_1 \, d\theta_2 - d\phi_1 \, d\phi_2 \, 
\sin\theta_1 \, \sin\theta_2)\bigr)\,. 
\end{split}
\]

In the limit $r \rightarrow \epsilon$, on surfaces $r^2 = \text{const.}$, 
the deformed conifold metric reduces to the $S^3$ surface metric.

We remark that the term $ds^2_3$ does not exist in the conifold metric \eqref{mc}.
The coordinate $\psi$ ceases to be a cyclic coordinate and only $\phi_1$ and $\phi_2$
continue to be cyclic. Therefore the number of the first integrals of the corresponding
Hamiltonian is insufficient to ensure the integrability of the geodesic flow.

\section{Conclusions}

In the last time it was proved the complete integrability of
geodesic motions on five-dimensional Sasaki-Einstein spaces $Y^{p,q}$ and $T^{1,1}$
\cite{BV,S-V-V,SVVAP}.

The purpose of this paper was to investigate the integrability of the 
metric cone of $T^{1,1}$ space and its resolved conifolds. We proved that the geodesics 
on the Calabi-Yau metric cone are completely integrable. This property is also 
valid for the small resolution of the conifold, but it is lost in the case of
deformation.

The formulation of the integrable systems in terms of action-angle variables permits
a comprehensive geometric description of their dynamics.

With reference to the metric cone over a Sasaki-Einstein space, the radial motion is 
unbounded and the corresponding Hamiltonian system, taken as a whole, does not admit 
a description with action-angle variables. Splitting the system into a ``radial'' and 
an ``angular'' part, the constants of motions are encoded in the ``angular 
Hamiltonian system''.

In view of fruitful extensions of the basic AdS/CFT correspondence, it is natural to 
expand the investigation of the integrability of geodesic flows for higher dimensional
Calabi-Yau spaces and their non-singular resolutions.

\section*{Acknowledgments}

This work has been partly supported by the joint Romanian-LIT,
JINR, Dubna Research Project, theme no. 05-6-1119-2014/2019 and partly
by the project NUCLEU 16 42 01 01/2016.


\begin{thebibliography}{99}

\bibitem{B-G-2008}
C. Boyer and K. Galicki,
{\it Sasakian geometry},
Oxford Mathematical Monographs, Oxford University Press, Oxford, 2008.

\bibitem{Sp}
J. Sparks, 
{\it Surv. Diff. Geom.} {\bf 16} (2011) 265.

\bibitem{JMM}
J. M. Maldacena,
{\it Adv. Theor. Math. Phys.} {\bf 2} (1998) 231.

\bibitem{GMSW}
J. P. Gauntlett, D. Martelli, J. Sparks and D. Waldram,
{\it Adv. Theor. Math. Phys.} {\bf 8} (2004) 711.

\bibitem{KW} 
I. R. Klebanov and  E. Witten,
{\it Nucl. Phys.} {\bf B536} (1988) 199.

\bibitem{BV}
E. M. Babalic and M. Visinescu,
{\it Mod. Phys. Lett. A} {\bf 30} (2015) 1550180.

\bibitem{MVEPJC}  
M. Visinescu,
{\it Eur. Phys. J. C} {\bf 76} (2016) 498.

\bibitem{MVPTEP}
M. Visinescu,
{\it Prog. Theor. Exp. Phys.} {\bf 2017} (2017) 013A01.

\bibitem{MP}
D. R. Morrison and M. R. Plesser
{\it Adv. Theor. Math. Phys.} {\bf 3} (1999) 1.

\bibitem{C-O} 
P. Candelas and X. de la Ossa, 
{\it Nucl. Phys.} {\bf B342} (1990) 246.

\bibitem{M-S}  
D. Martelli and J. Sparks, 
{\it Commun. Math. Phys.} {\bf 262} (2006) 51.

\bibitem{S-V-V}
V. Slesar, M. Visinescu and G. E. V\^ilcu, 
{\it EPL} {\bf 110} (2015) 31001.

\bibitem{FMW}
R. Field, I. V. Melnikov and B. Weaver,
{\it Dynamics on asymptotically conical geometries},
arXiv:1710.00404 [hep-th] (2017).

\bibitem{HKLN}
T. Hakobyan, S. Krivonos, O. Lechtenfeld and A. Nersessian,
{\it Phys. Lett. A} {\bf 374} (2010) 801.

\bibitem{LNY}
O. Lechtenfeld, A. Nersessian and V. Yeghikyan,
{\it Phys. Lett. A} {\bf 374} (2010) 4647.

\bibitem{HLNSY}
T. Hakobyan, O. Lechtenfeld, A. Nersessian, A. Saghatelian and V. Yeghikyan,
{\it Phys. Lett. A} {\bf 376} (2012) 679.

\bibitem{GPS}
H. Goldstein, C. Poole and J. Safko,
{\it Classical Mechanics}, 3rd edition
(Addison-Wesley, San Francisco, 2002).

\bibitem{PZT}
L. A. Pando Zayas and A. A. Tseytlin, 
{\it JHEP} {\bf 11} (2000) 028.

\bibitem{OY}
K. Ohta and T. Yokono, 
{\it JHEP} {\bf 02} (2000) 023.

\bibitem{MT}
R. Minasian and D. Tsimpis,
{\it Nucl. Phys.} {\bf B572} (2000) 499.

\bibitem{SVVAP}
V. Slesar, M. Visinescu and G. E. V\^ilcu,
{\it Annals Phys.} {\bf 361} (2015) 548.

\end{thebibliography}
\end{document}